\documentstyle[psfig,emulatepasa]{article}
\def\reference{\parskip 0pt\par\noindent\hangindent 0.5 truecm}

\def\spose#1{\hbox to 0pt{#1\hss}}
\def\simlt{\mathrel{\spose{\lower 3pt\hbox{$\mathchar"218$}}
     \raise 2.0pt\hbox{$\mathchar"13C$}}}
\def\simgt{\mathrel{\spose{\lower 3pt\hbox{$\mathchar"218$}}
     \raise 2.0pt\hbox{$\mathchar"13E$}}}

\def\OVI{O{\small\sc{VI}~}}

\begin{document}

\small
\shorttitle{Is High-Velocity Cloud Complex~C Associated with the Galactic Warp?}
\shortauthor{D.\ Kawata, C.\ Thom, \& B.K.\ Gibson}
%
%
\title{\large \bf
Is High-Velocity Cloud Complex~C Associated with the Galactic Warp?}

\author{\small 
 Daisuke Kawata$^{1}$,
 Christopher Thom$^{2}$,
 Brad K.\ Gibson$^{3}$ 
} 

\date{}
\twocolumn[
\maketitle
\vspace{-20pt}
\small
{\center
Centre for Astrophysics \& Supercomputing, 
Swinburne University, Mail \#31, P.O. Box 218, Hawthorn, VIC 3122, 
Australia\\
$^1$dkawata@astro.swin.edu.au, $^2$cthom@astro.swin.edu.au,
$^3$bgibson@astro.swin.edu.au\\[3mm]
}

%
\begin{center}
{\bfseries Abstract}
\end{center}
\begin{quotation}
\begin{small}
\vspace{-5pt}
    We test the hypothesis that High-Velocity gas cloud Complex~C is actually
    a high-latitude spiral arm extension in the direction of the Galactic
    warp, as opposed to the standard interpretation - that of a once
    extragalactic, but now infalling, gas cloud.  A parallel Tree
    N-body code was employed to simulate the tidal interaction of a
    satellite perturber with the Milky Way.  We find that a model
    incorporating a perturber of the mass of the Large Magellanic Cloud on a
    south-to-north polar orbit, crossing the disk at $\sim$15~kpc, does yield a
    high-velocity, high-latitude, extension consistent with the spatial,
    kinematical and column density properties of Complex~C.  Unless
    this massive satellite remains undiscovered because of either a
    fortuitous alignment with the Galactic bulge (feasible within the
    framework of the model), or the lack of any associated baryonic
    component, we conclude that this alternative interpretation appears
    unlikely.
\end{small}
\end{quotation}
]


\bigskip

\section{Introduction}

Despite 40 years of intense study, the origin of thousands of High-Velocity
Clouds (HVCs) remains largely a mystery. HVCs are clouds of hydrogen gas
that exhibit velocities that do not conform to a simple model of galactic
rotation.  Covering more than a third of the sky at radio wavelengths
(Murphy, Lockman \& Savage 1995)\footnote{Perhaps as much as 85\% of sky,
  as inferred by probes of lower column-density \OVI gas (Sembach et~al.
  2003).}, their origin scenarios fall into two broad categories (see
Wakker \& van~Woerden 1997 for a comprehensive review) - Galactic and
extragalactic.

Under the Galactic scenario, HVCs are considered to have a ``local''
origin, perhaps being infalling debris initially driven from the disk into
the halo by a supernova- or stellar wind-driven ``galactic fountain''
(Bregman 1980), or perhaps being simply a high-latitude extension of the
outer spiral arms due to the warp of the disk (Davies 1972).

The extragalactic scenario encompasses several disparate suggestions
ranging from the tidal disruption of gas-rich satellites such as the
Magellanic Clouds (Murai \& Fujimoto 1980; Gardiner \& Noguchi 1996), to
the infall of proto-galactic Cold Dark Matter (CDM) Local Group building
blocks (Blitz et~al. 1999).  In these pictures, HVCs may represent the
infalling metal-poor gas invoked to reproduce the metallicity distribution
function of the solar neighbourhood by many chemical evolution models of
the Milky Way (Twarog 1980; Gibson et~al. 2002).  The most extreme variants
of these extragalactic hypotheses (e.g. Blitz et~al.) suggest that HVCs
are in fact the baryonic tracer of the dark matter satellites predicted to
exist (by cosmological CDM simulations) in large numbers throughout the
Local Group (Klypin et~al. 1999; Moore et~al. 1999).

In recent years the HVC which has come under the greatest scrutiny is
Complex~C. This cloud covers over 1600 deg$^2$ at high Galactic latitudes
($+$20$^\circ$ $\simlt$ $b$ $\simlt$ $+$60$^\circ$), in quadrants I and II
of the Milky Way.  While explicitly excluded from the Blitz et~al.  (1999)
``extragalactic origin'' model, Wakker et~al. (1999a) suggest that
Complex~C is in fact an infalling extragalactic cloud, based upon its
apparently low metallicity (c.f. Gibson et~al.  2001).  Such an
interpretation opens up the possibility that Complex~C may be the baryonic
component of one of the numerous halos predicted by the cosmological CDM
simulations referred to earlier, which in turn could lend support to the
notion that many such HVCs trace cosmologically ``interesting'' structures
(as opposed to being more of interest in a Galactic sense). Recently, Tripp
et~al. (2003) have obtained a high resolution {\it STIS} spectrum of
3C~351, in which they find evidence for the hypothesis that Complex~C is
ablating and dissipating as it approaches the plane of the Galaxy.  This
idea is supported by {\it FUSE} \OVI observations (Wakker et~al, 2003;
Sembach et~al, 2003).

Converse to this picture, as early as 30 years ago Davies (1972)
recognised that Complex~C had a surprising projected spatial and
kinematical connection to the Galaxy in the direction of the disk's warp in
Quadrant~I.  Davies' hypothesis placed Complex~C in the outer Milky Way,
either as a high-latitude extension of the Outer Arm (where the maximum in
the disk's warp occurs), or as a separate high-latitude spiral arm.  Davies
suggested that the passage of the Large Magellanic Cloud on its orbit may
have been responsible for both the warp and high-latitude extension
(comprising Complex~C), albeit without the benefit of testing that
hypothesis computationally.

In what follows, we revisit this alternative suggestion as to the origin of
Complex~C - specifically, \it can the interaction of a satellite galaxy
with the Milky Way induce a high-latitude extension of the Outer Arm
consistent with the spatial and kinematical characteristics of Complex~C?
\rm If such a model can be constructed, \it what are the implications for
the mass and orbit of the satellite?\rm and \it what is the likelihood that
the model reflects reality? \rm To address these three questions, we employ
a series of Tree N-body simulations over a wide range of mass- and
orbital-parameter space.  In Section~2 we describe the numerical framework
adopted to simulate the encounter of a satellite galaxy with that of the
Milky Way.  The results of these simulations are then presented in
Section~3, with the accompanying discussion and conclusions provided in
Section~4.

\section{Model}

The numerical simulations modelling the interaction of the Milky Way and a
companion dwarf galaxy were conducted using a Tree N-body code.  This code
forms part of GCD+, our original parallel Tree N-body/SPH code described in
Kawata(1999) and Kawata \& Gibson (2003). The Milky Way (consisting of live
halo, disk and bulge) was constructed using the MW-A model of Kuijken \&
Dubinski (1995). We used a disk scale-length of $R_D=4.5$~kpc and rotation
velocity of $V_{\rm rot}=220$~kms$^{-1}$.  The halo, disk and bulge were
described using 200,000, 80,000 and 20,000 particles respectively, giving
particle masses of $1.24\times10^6$~M$_\odot$ (halo),
$1.08\times10^6$~M$_\odot$ (disk) and $5.51\times10^5 $~M$_\odot$ (bulge).
The softening lengths (which are proportional to the particle mass to the
1/3rd power) of the halo, bulge and disk were set to 0.37~kpc, 0.35~kpc and
0.28~kpc respectively.  Throughout the paper, we consider north to be along
the z-axis, with the disk in the $x$-$y$ plane.  The rotation axis of the
Milky Way is set to the negative direction of $z$-axis. Under these
conditions, we have confirmed the stability of the Milky Way model to the
same degree as that shown by Kuijken \& Dubinski (1995).

For simplicity, the dwarf companion is modelled by a rigid Plummer sphere
of scale-length 1~kpc.  We present three simulations showing how close an
orbit, and how massive a companion, are required to reproduce high latitude
gas clouds like Complex~C as an extension of the Galactic Warp. The initial
position, velocity and mass of the perturber for each model are shown in
Table~\ref{tbmodelp}. Here, model A is our fiducial model.  Model B has a
reduced mass for the perturber, while model C is given a larger radius
orbit. Once the initial position and velocity of the perturber are
specified, the orbit of the perturber is self-consistently calculated by
the N-body scheme.  Since we construct both halo and disk with live N-body
particles (as opposed to assuming a fixed potential), the effects of
dynamical friction on the orbit are naturally taken into
account\footnote{The effect of the dynamical friction is mainly determined
  by the mass of the perturber and the background density, and does not
  depend on the Plummer scale-length (Noguchi, 1999).}

Complex~C is seen in the Northern sky and we wish to consider the case
where Complex~C is a high scale-height extension of the Outer Arm, from the
disk towards the North Galactic Pole. The orbit required to create such an
extension should pass from south to north through the disk.  We assume this
orbit in all models. This is a simple and idealised condition.  We are
primarily interested in tidal forces in the z-direction and therefore only
consider the (simplified) case of polar orbits.  All the known close
satellites of the Milky Way (i.e., LMC, SMC, Sagittarius dwarf) occupy
polar orbits. This restriction is therefore useful in limiting the
parameter space to be explored and enabling us to compare to the observed
Milky Way satellites, whilst not limiting the usefulness of the model

We focus only on the effect of a single passage of the perturber.  It would
also be interesting to observe the effects of many passages of the
satellite.  To follow such a long evolution, however, requires that the
mass-loss of the companion be taken into account.  For this idealised
scenario, we wish to limit the parameter space to be explored and thus only
follow a single passage of the perturber.  The simulations were run for
0.33~Gyr, allowing sufficient time for this single interaction to occur.

In this paper, we model the components of the Milky Way using collisionless
particles.  We thus only follow the effects of gravity on the dynamical
evolution of the system.  We recognise that both the HVC and outer disk
(the region we are primarily interested in) are dominated by gas and we
should, in principle, consider hydrodynamical effects.  Since the gas
densities are low, however, we may neglect collisional effects in the gas
components for this initial study.

\begin{table*}[tp]
\caption{Initial Parameters for Perturber }
\label{tbmodelp}
\begin{center}
\begin{tabular}{clcc}
  \hline
  Model  &  Mass (M$_\odot$) & Position ($x,y,z$~kpc) 
  & Velocity ($v_x,v_y,v_z$ kms$^{\rm -1}$)\\
  \hline
  A & $2\times10^{10}$ & $-20, 0, -20$ & 0, 0, 187    \\
  B & $2\times10^{9}$ & $-20, 0, -20$ & 0, 0, 187    \\
  C & $2\times10^{10}$ & $-40, 0, -20$ & 0, 0, 187    \\
  \hline
 \end{tabular}
\end{center}
\end{table*}

\section{Results}

\begin{figure*}[t]
 \begin{center}
 \begin{tabular}{ccc}
   \psfig{file=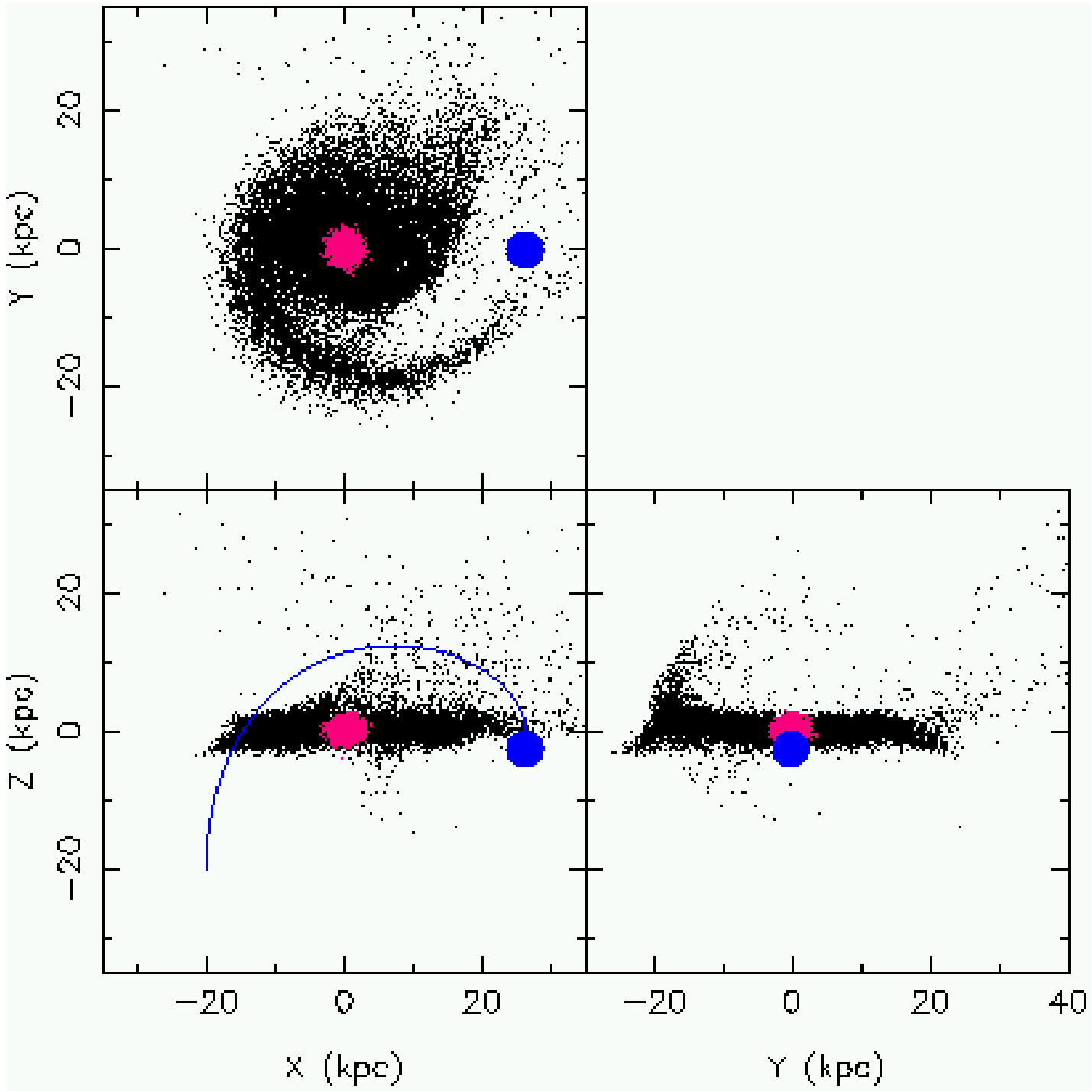,width=8.0cm} &
   \psfig{file=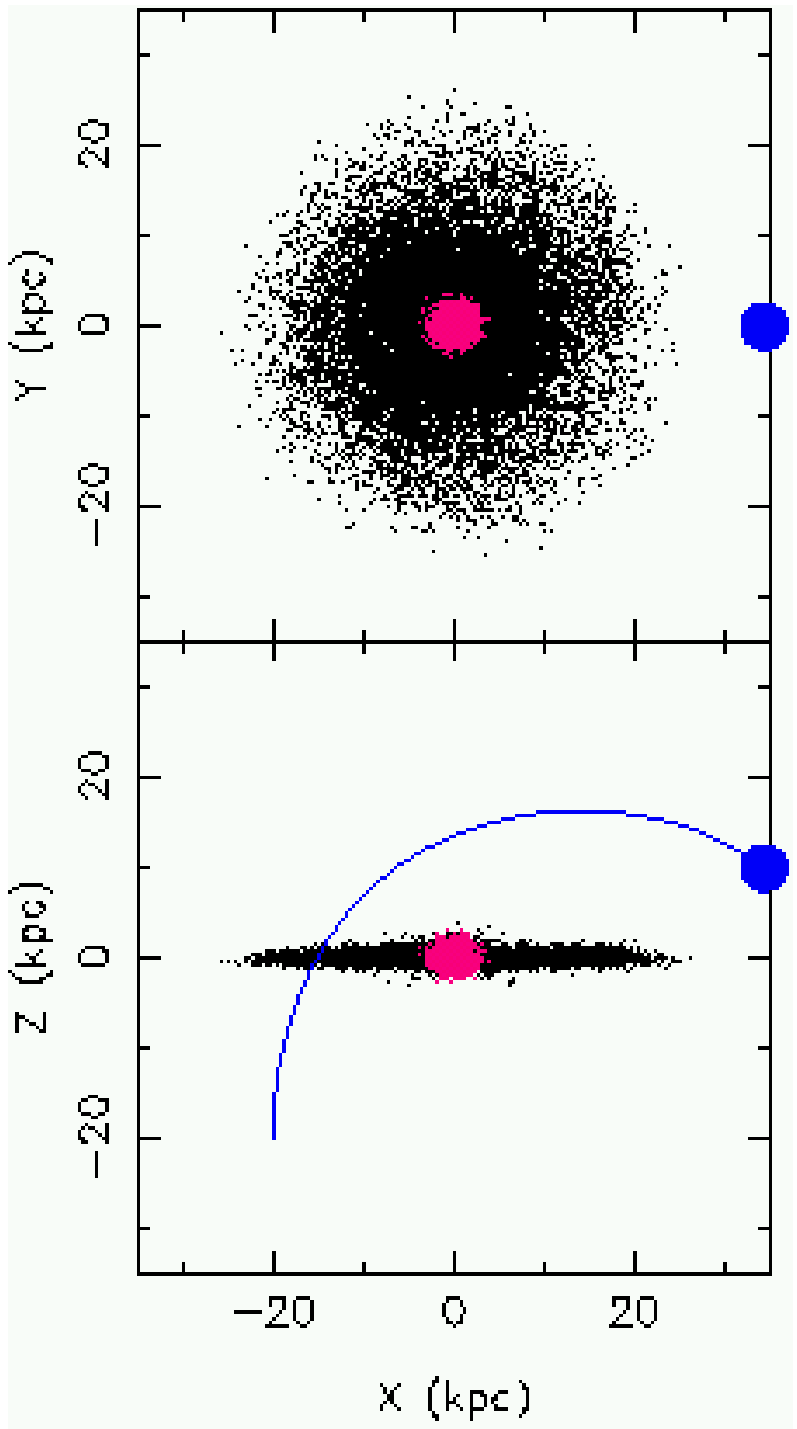,width=4.0cm} &
   \psfig{file=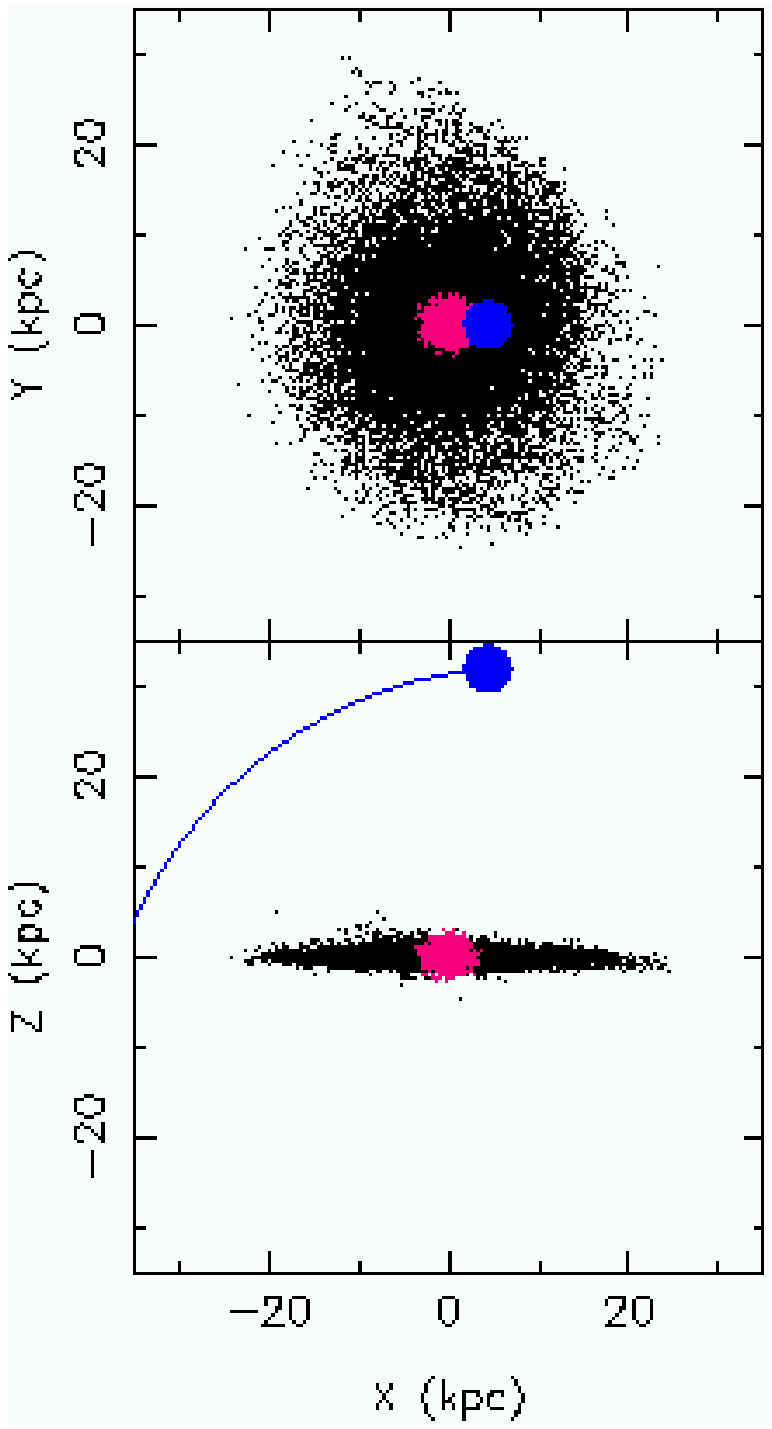,width=4.0cm} \\
 \end{tabular}
\caption{ Final particle distribution of models A (upper), B (lower left),
  and C (lower right). The solid line shows the orbit of the perturber
  (shown in blue).  Only the disk (black dots) and bulge (red dots)
  particles are plotted.  The rotation axis of the Milky Way is the
  negative direction of the z-axis (into the page on the $x-y$ plots). The
  disk is rotating clockwise in the upper panels. Since model A is not
  axi-symmetric (due to the extended warp), three different projections are
  presented. Models B and C are close to axi-symmetric and hence only two
  panels are shown. In the y-z projection for model A, two warps may be
  seen extending north out of the disk.  For the two sub-models A1 and A2,
  we assume the position of the sun to be $(x,y) = (-8.5,0)$ and $(8.5,0)$
  respectively.  In sub-model A1, the right-hand warp corresponds to the
  position of Complex~C (with the left hand region being the Complex~C
  position for sub-model A2).}
 \label{flast}            
 \end{center}
\end{figure*}

Figure~\ref{flast} shows the final particle distributions of all models.
As can be seen, model A is the only one in which the perturbing dwarf
affects the disk.  In the model A simulation, two spiral-arm patterns are
induced by the tidal force of the perturber and some particles are drawn
out of the disk along the orbit of the satellite.  As shown below in
detail, model A provides the best set of parameters for reproducing gas
clouds with similar position, kinematics and column densities to Complex~C.
From this, we infer that to create the extended warp the perturber must be
as massive as the Large Magellanic Cloud (LMC) ($\sim$2$\times10^{10}$
M$_\odot$) and cross the disk on a south-to-north orbit as close as 15~kpc
from the Galactic centre.  We now focus on the results for this model.

\begin{table}[htp]
\caption{Assumed position and velocity of the sun in each sub-model}
\label{tbpvsun}
\begin{center}
\begin{tabular}{ccccc}
\hline
 & \multicolumn{2}{c}{Position (kpc)} &  
   \multicolumn{2}{c}{Velocity (kms$^{-1}$)} \\
 Model & $x$ & $y$ & $v_x$ & $v_y$ \\
\hline
  A1 & $-8.5$ & 0 & 0 & -220 \\
  A2 & 8.5 & 0 & 0 & 220 \\
 \hline
 \end{tabular}
\end{center}
\end{table}
In the $y-z$ projection in the upper panel of Figure~\ref{flast}
we see two warps bending northwards out of the disk.  We
consider two sub-models (A1 and A2), in which we assume a solar position
such that each warp would be located at a similar Galactic longitude to Complex~C.
Table~\ref{tbpvsun} shows the assumed solar position and velocity in these
sub-models. For simplicity, we ignore the slight difference between the
solar velocity and the velocity of the Local Standard of Rest, choosing the
LSR value.

\begin{figure*}[t]
 \begin{center}
   \begin{tabular}{cc}
     \psfig{file=lb0-v.ps,width=8.25cm} &
     \psfig{file=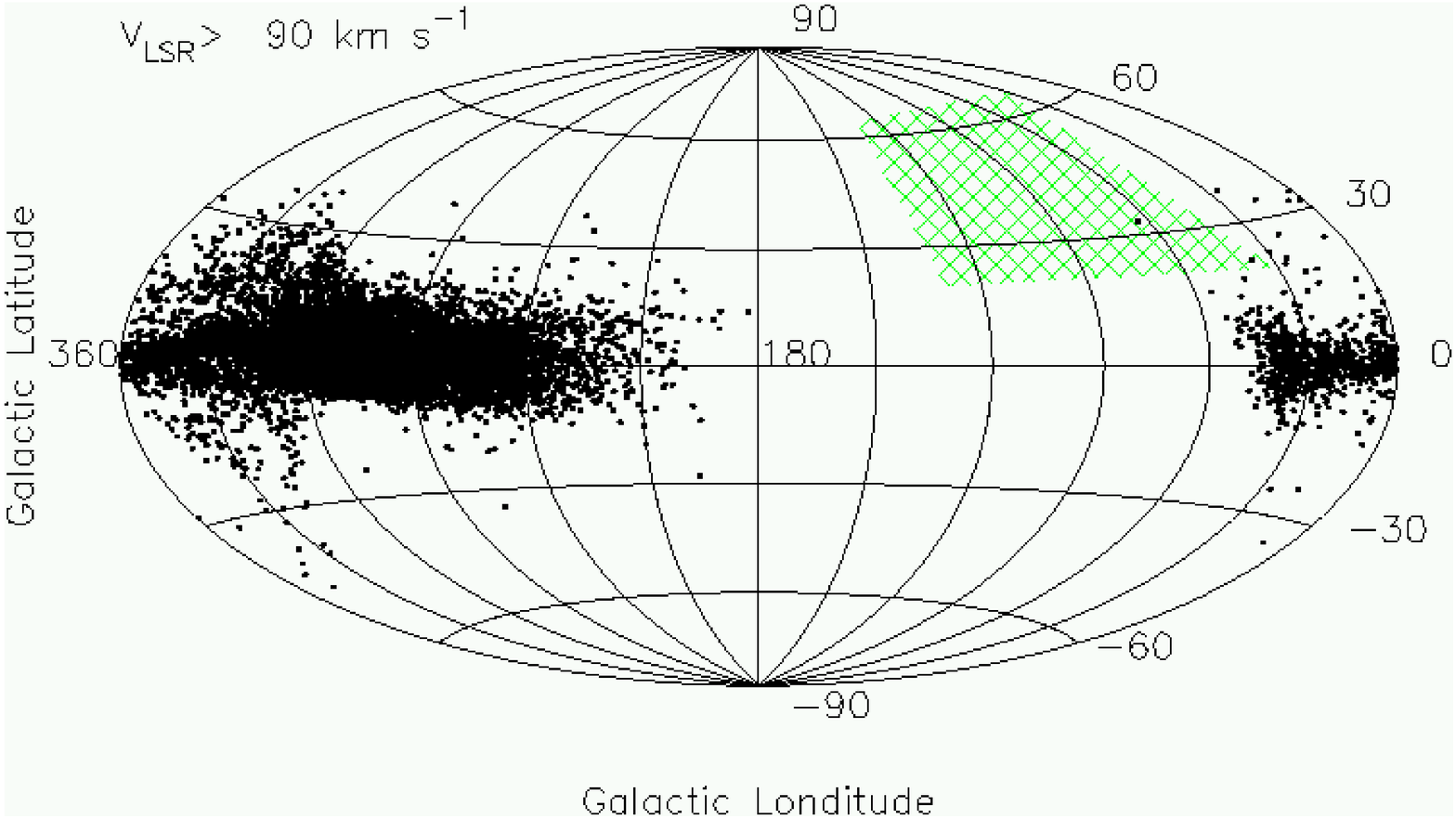,width=8.25cm} \\
   \end{tabular}
   \caption{ Final particle distribution in the $l-b$ plane for sub-model A1.
     The two panels show all particles with $V_{LSR}<-90$~kms$^{-1}$ (left
     panel) and $V_{LSR}>90$~kms$^{-1}$ (right panel). The cross-hatched
     region indicates the position of Complex~C.}
 \label{flb0v6}            
 \end{center}
\end{figure*}

\begin{figure*}[t]
 \begin{center}
   \begin{tabular}{cc}
     \psfig{file=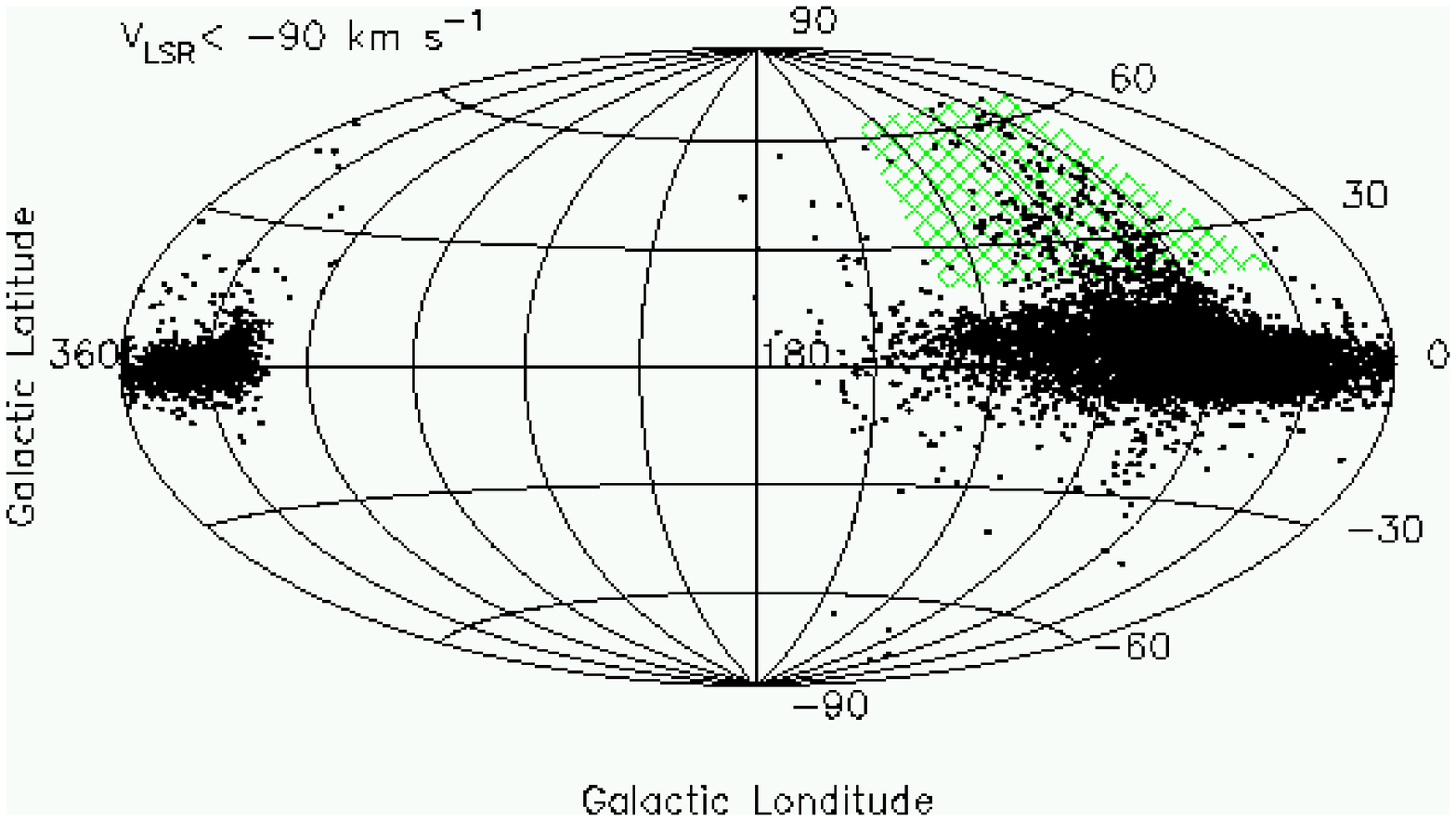,width=8.25cm} &
     \psfig{file=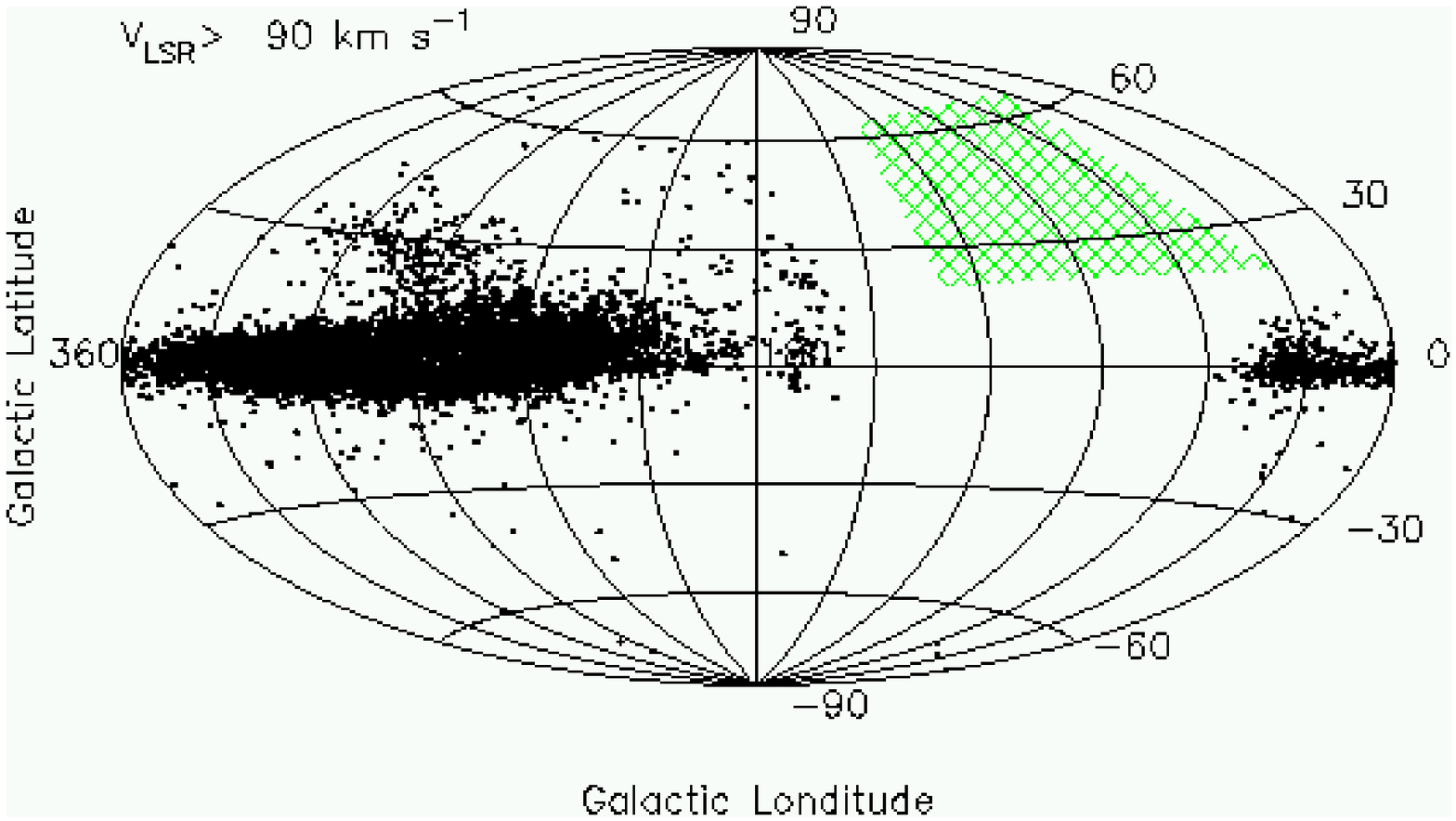,width=8.25cm} \\
   \end{tabular}
   \caption{ Same as Figure~\ref{flb0v6}, but for sub-model A2.}
   \label{flb180v6}            
 \end{center}
\end{figure*}

Figures~\ref{flb0v6} \& \ref{flb180v6} show all-sky positive- and
negative-velocity projections of the final particle distributions for the
two sub-models.  In both figures, particles with galactocentric radius
$R<5$~kpc have not been plotted, since we are interested only in the outer
region.  The position of Complex~C in Galactic coordinates is roughly
($l=50 - 150^{\circ}$, $b=20 - 65^{\circ}$) with a mean velocity of
V$_{LSR}=-150\sim-110$~kms$^{-1}$ (Wakker 2001).  In both sub-models, a
significant number of particles can be seen in this area, with similar
velocities to that of Complex~C.  Note that in both images, only particles
with negative velocities populate the position of Complex~C.

We also note that sub-model A2 includes a fortuitous group of particles
with positive velocities in quadrant III/IV of the galaxy.  This region
corresponds to the same area of sky as population EP and HVC complexes WA,
WB and WD. This component comes from the high-latitude extension of the
warp, which is seen in the Complex~C region in sub-model A1 (red circles in
Figure 4).  This is an interesting coincidence and natural outcome of the
model.  Population EP has higher V$_{LSR}$ ($> 200$ kms$^{-1}$) than
complexes WA/WB/WD ($< 200$ kms$^{-1}$).  The mean velocity of particles
in the region ($l=210 - 300^{\circ}$, $b=15 - 60^{\circ}$) is
V$_{LSR}\sim207$~kms$^{-1}$.  Although the spatial resolution makes it
difficult to decided whether to associate Population~EP or
Complexes~WA/WB/WD with the group of particles, the position and velocity
are suggestive of Population~EP. Population EP, or at least some fraction
of it, has been suggested to correspond to gas associated with the Leading
Arm of the Magellanic Stream (Gardiner \& Noguchi 1996; Putman
et~al. 1998), or a gas filament falling into the Local Group (Blitz
et~al. 1999).  The warp origin is a possible alternative explanation for
population EP. Distance measurements for population EP would be crucial
test for these scenarios.

\begin{figure*}[t]
  \begin{center}
    \psfig{file=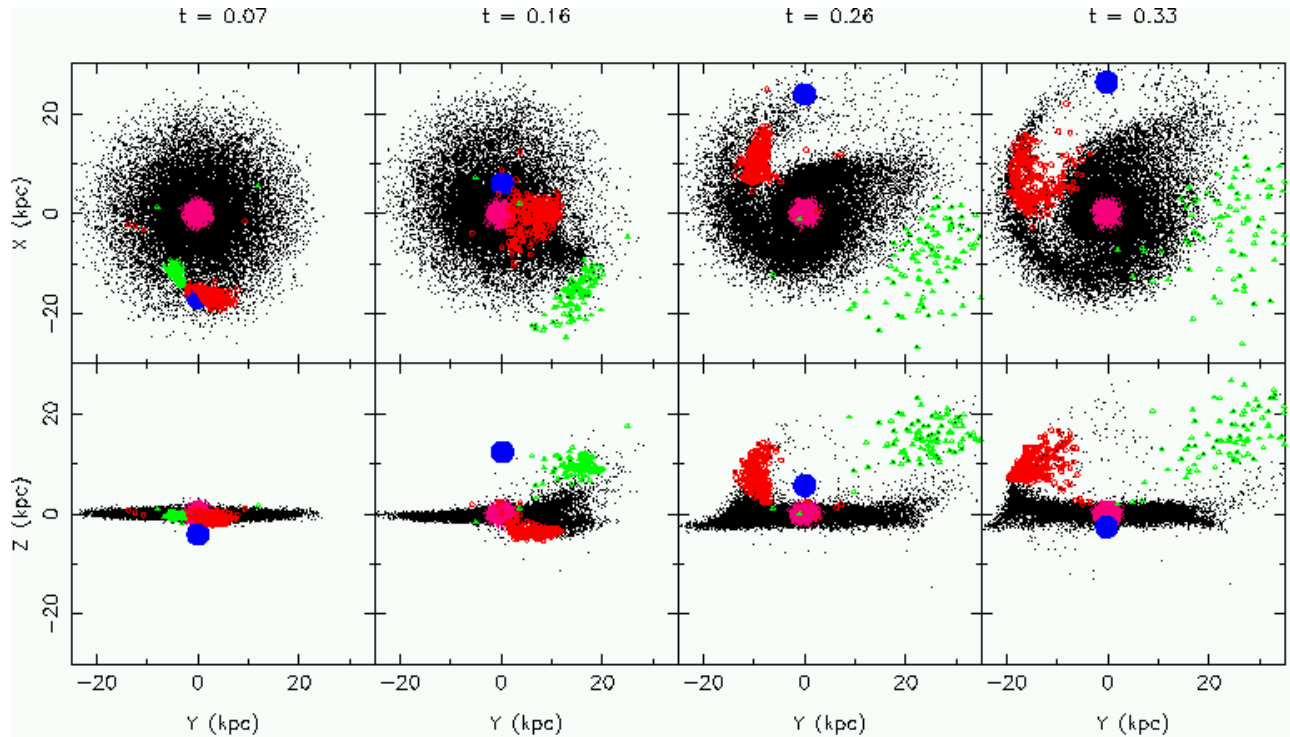,width=17cm}
    \caption{ Morphological Evolution of model A. Only disk (black dots)
      and bulge (grey dots) particles are plotted. Dots enclosed by small
      triangles (in the case of sub-model A1) or circles (for sub-model A2)
      and are the particles which settle in region of Complex~C.  The solid
      circle represents the position of the perturber.  The rotation axis
      of the Milky Way is the negative direction of the z-axis.  The upper
      panels show the Milky Way as seen from the South Galactic Pole (and
      hence the disk is rotating counter-clockwise). The units of time
      shown at the top of the panels are Gyr.}
    \label{fm2e10x2anim}            
  \end{center}
\end{figure*}

To date, no stellar content associated with Complex~C has been detected.
We would therefore expect any particles in the region of Complex~C to have
come from a gas-dominated region i.e., the edge of the disk.
Figure~\ref{fm2e10x2anim} shows the morphological evolution of model A,
with the perturber crossing the disk at a radius of $\sim$15~kpc.  Disk
particles are drawn up out of the disk by the tidal force.  The face-on
view shows the spiral pattern induced by the tidal interaction.  The
particles which finally settle in the Complex~C region have been tagged by
triangles (for sub-model A1) or circles (sub-model A2).  Before the
interaction, these particles lie at the edge of the disk (t=0.07 of
Figure~\ref{fm2e10x2anim}).  We thus justify the claim that, at the end of
the simulation, particles in the region of Complex~C should represent
mainly gas, in agreement with observations.  

We also note that, since the particles originate from well outside the
solar-circle, the metallicity of the gas would be expected to be sub-solar.
Complex~C has been extensively studied in the UV, in order to obtain
metallicity information (e.g. Wakker et~al, 1999a; Gibson et~al, 2001;
Tripp et~al, 2003).  The reported metallicities have differed, depending on
the sight-lies studied and techniques used to derive abundances.  Recent
studies (Tripp et~al, 2003; Collins et~al, 2003) agree on a metallicity
range of $Z\sim0.1-0.3$~Z$_{\odot}$. Under the conditions we present,
drawing gas from the disk at a radius of 15~kpc, we would expect
metallicities consistent with this range based on observations of disk gas
metallicities (e.g., Afflerbach et al. 1997).

It is worth noting that, just before the perturber crosses the disk,
particles represented by circles (see Figure~\ref{fm2e10x2anim}) lead the
perturber (i.e. are ahead in the disk rotation direction), whilst those
particles shown as triangles trail the perturber.  The evolution of the two
sets of particles is quite different.  In the case of the triangles, the
perturber's tidal force increases the angular momentum of the particles.
This increase in angular momentum throws the particles to the outer disk as
they slowly move up out of the disk (towards galactic north).  Conversely,
the particles leading the perturber will have a net decrease in angular
momentum.  This causes the particles to fall in towards the bulge.  Since
the perturber is initially on the south side of the disk, these particles
first move downward, before they are dragged north of the disk as the
perturber moves on its orbit up out of the disk.  Thus the morphology of
the two sets of particles is determined by their original position with
respect to the satellite.

\begin{figure*}
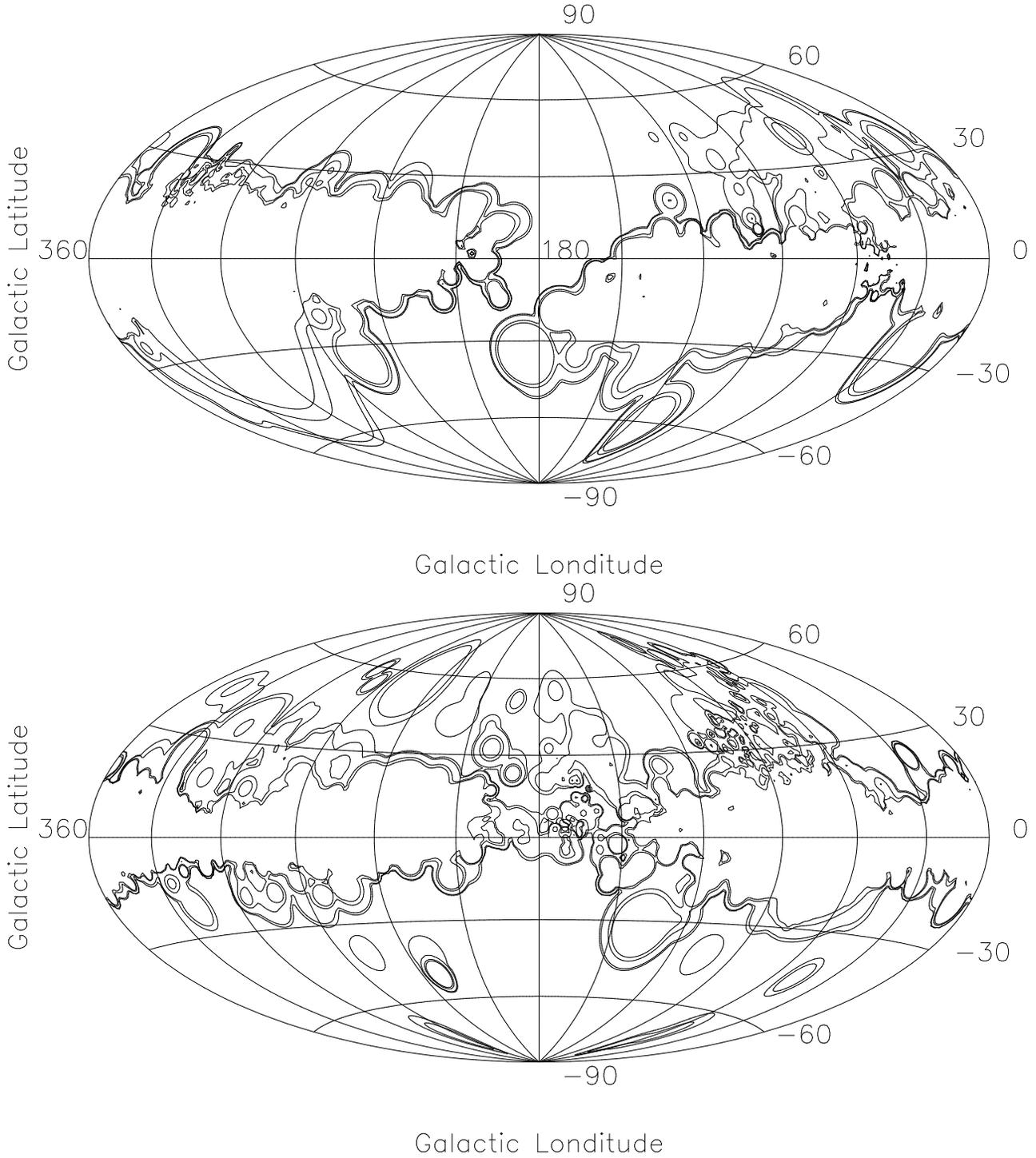

 \begin{center}
 \begin{tabular}{c}
  \psfig{file=ncollb0.ps,width=17cm} \\
  \psfig{file=ncollb180.ps,width=17cm} 
 \end{tabular}
 \caption{ Column density map of model A1 (upper) and A2 (lower). 
Contours are drawn at column density levels $2\times 10^{18}$,
$5\times 10^{18}$, $2\times 10^{19}$ and $10^{20}$~cm$^{-2}$.}
 \label{fncollb}            
 \end{center}
\end{figure*}

A final consideration is the (HI) column density.  Figure~\ref{fncollb}
shows the column density contour map for each sub-model.  The plotted
column density is the projected density of all high-velocity particles
($|V_{\rm LSR}|>100$~kms$^{-1}$ - this level, and the contour levels, are
the same as those used in Figure~2 of Wakker (1991)), assuming that 76$\%$
of the mass of each particle is neutral hydrogen (this is the primordial
value and hence the largest possible fraction).  Again, particles with
Galactocentric radius less than 5~kpc have been excluded.  As Wakker (1991)
notes, the column density of Complex~C is between
$2\times10^{18}$~cm$^{-2}$ and $5\times10^{18}$~cm$^{-2}$, and more than
$5\times10^{18}$~cm$^{-2}$ in some small parts.

Figure~\ref{fncollb} demonstrates that both models A1 and A2 successfully
reproduce an extended HI gas structure in the region of Complex~C. The
predicted HI column density is slightly higher than that of Complex~C, but
this prediction will be an upper limit on the value.  If the metallicity of
the gas is not primordial, and/or some fraction of the hydrogen is ionised,
then we expect the (predicted) column density to decrease\footnote{The
  assumption of no ionisation gives a conservative upper limit.  In
  practice, H$\rm \alpha$ measurements show the ionisation factor is
  non-zero (Wakker, 1999a; Wakker, 1999b). }. Hence the predicted column
density for both models acceptably agrees with the observed Complex C
column density. Since the HI fraction of Complex~C is not well-known,
further precise comparison is difficult.

Table~\ref{tbpcc} summarises the properties of the Complex~C candidates in
sub-models A1 and A2, which are measured by counting the particles with
$V_{\rm LSR}<-90$kms$^{-1}$ in the region ($l=50\sim150^{\circ}$,
$b=20\sim65^{\circ}$).  As also seen in Figures~\ref{flb0v6} and
\ref{flb180v6}, both models succeed in reproducing the kinematical
characteristics of Complex~C. Neither the total mass of, nor distance to,
Complex~C are known. The column density discussed above is a combination of
these two factors.  Non-detections of CaII absorption by Complex~C in 5
stellar probes have set a firm lower limit of 1.2~kpc on the distance to
Complex~C (discussed by Wakker (2001)).  A less firm limit of 6.1~kpc is
set by further CaII non-detections.  These limits are not particularly
restrictive and not inconsistent with the model presented here.

\begin{table}
\caption{Properties of Complex~C candidates in models}
\label{tbpcc}
\begin{center}
\begin{tabular}{cccc}
\hline
       & Total mass &  Mean velocity  & Distance \\
 Model & (M$_{\odot}$) & (kms$^{-1}$) & (kpc)  \\
\hline
  A1 & $5.4\times10^7$ & $-145$ & 32.7 \\
  A2 & $1.6\times10^8$ & $-140$ & 18.4 \\
 \hline
 \end{tabular}
\end{center}
\end{table}

\section{Discussion and Conclusion}
Using a series of high-resolution N-body simulations, we modelled the
interaction of the Milky Way (consisting of bulge, disk and live halo) with
a dwarf companion galaxy.  We found that a perturber with mass of order
$2\times10^{10}$~M$_\odot$ (a similar size to the LMC) and which crosses
the disk at a Galactocentric radius of order 15~kpc can induce a
high-latitude extension of the Outer Arm.  This extension is consistent
with the spatial, kinematical and column-density characteristics of
Complex~C. We have also shown in Figure~\ref{flast} that such a
high-latitude extension cannot be induced by a perturber with mass smaller
than $2\times 10^{10}$~M$_\odot$ or with an orbit which crosses the disk at
greater than $\sim$15~kpc.

According to recent orbital analysis of the LMC (e.g. Gardiner \& Noguchi
1996), the pericentre of the LMCs orbit is 45~kpc and thus it must cross the
disk plane well outside the 15~kpc boundary. Our study therefore suggests
that the LMC {\it cannot} be responsible for such a high scale-height
extension of the Outer Arm.  This is consistent with the prediction of
Hunter \& Toomre (1969). 

The predicted orbit of the recently discovered Sagittarius dwarf galaxy has
a smaller perigalacticon of $\sim$15~kpc (Ibata \& Lewis 1998). Despite its
favourable orbit, however, the mass of the Sagittarius dwarf is smaller
than the LMC by an order of magnitude ($\sim$10$^9$ M$_\odot$) and the
predicted last passage through the disk is from North to South at a radius
of $\sim$70~kpc.

Thus, our scenario would require a large, heretofore undiscovered,
companion dwarf galaxy. It may be possible for such a perturber to be
obscured by the Galactic Bulge.  Figure~\ref{fm2e10x2anim} shows this is
possible within the confines of our simulation (sub-model A1). If such a
satellite had such a close orbit, a stellar and/or gas tidal stream would
be expected.  The formation of such a stream is, however, dependant on the
stellar/gas mass to dark halo mass ratio (Johnston et al., 1996; Ibata et
al., 2001)\footnote{Also, a luminous, baryonic stream is only expected if
  the perturber initially has such a component. A pure Dark Matter halo
  would leave no such stream}.  To date, there is no observational evidence
for such a stream.  Another possibility is the lack of any associated
detectable baryonic component, i.e., a dark galaxy (e.g., Trentham et~al.
2001).  We view both possibilities as highly unlikely. Further,
Figures~\ref{flast} and \ref{fm2e10x2anim} show that a satellite with the
required mass and orbit to create Complex~C would be massive enough to
significantly thicken the thin disk. We have measured exponential
scale-heights of the disk and obtained $\sim0.3$~kpc before the interaction
and $\sim0.5$~kpc after the interaction. The final disk thickness is
inconsistent with observations of the scale-height of the thin disk
(0.3~kpc).  We thus take our results to be strong (but not conclusive)
evidence against the suggestion of Davies (1972) that Complex~C is a high
scale-height extension of the Outer Arm induced by a past interaction with
a satellite dwarf galaxy.

A dwarf satellite interaction is not the only way to make such an extension
of the Outer Arm. Recently, Bekki \& Freeman (2002) showed that a rotating
triaxial dark halo can induce a high-latitude extension of the outer disk
gas. This model may be able to explain the scenario in which Complex~C is
part of the Outer Arm, although according to their paper, the
quantitative comparison of the kinematical and spatial properties of
Complex~C with numerical simulation is still in progress.

The present study provides no discussion on the validity of other formation
scenarios for Complex~C (e.g. the Galactic Fountain or extragalactic
models).  Complex~C is the largest HVC and its kinematical and chemical
properties have been extensively observed using various facilities from the
UV to radio wavelengths. This paper has shown that the observed kinematical
properties give useful constraints on one formation scenario.  To
understand the origin of Complex~C, as well as other HVCs, quantitative
comparisons between observation and numerical simulations (both chemical
and dynamical properties) are important. In future work, we plan to test
other formation models, taking advantage of our original chemo-dynamical
evolution code, GCD+.

\section*{Acknowledgements}

We acknowledge the Yukawa Institute Computer Facility, the Astronomical
Data Analysis Centre of the National Astronomical Observatory Japan, and
the Victorian and Australian Partnerships for Advanced Computing, where the
simulations described here were performed.  This work is supported by the
Australian Research Council (A0010517 and DP0343508) and the Swinburne
University Research Development Grants Scheme.

\section*{References}
\reference Afflerbach, A., Churchwell, E. \& Werner, M.W.,
 1997, ApJ, 478, 190
\reference Bekki, K. \& Freeman, K.C. 2002, ApJ, 574, L21
\reference Blitz, L., Spergel, D.N., Teuben, P., Hartmann, D. \&
 Burton, W.B., 1999, ApJ, 514, 818
\reference Bregman, J.N., 1980, ApJ, 236, 577
\reference Collins, J.A., Shull, J.M., \& Giroux, M.L., 2003, ApJ, 585, 336 
\reference Davies, R.D., 1972, MNRAS, 160, 381
\reference Gardiner, L. \& Noguchi, M., 1996, MNRAS, 278, 191
\reference Gibson, B.K., Fenner, Y., Maddison, S.T. \&
 Kawata, D., 2002, in Extragalactic Gas at Low Redshift, eds.
 J.S. Mulchaey \& J.T. Stocke, (San Francisco: ASP), 225
\reference Gibson, B.K., Giroux, M.L., Penton, S.V.,
 Stocke, J.T., Shull, M. . \& Tumlinson, J.,
 2001, AJ, 122, 3280
\reference Hunter, C., \& Toomre, A., 1969, ApJ, 155, 747
\reference Ibata, R.A., \& Lewis, G.F., 1998, ApJ, 500, 575
\reference Ibata, R.A., Irwin, M., Lewis, G.~F., \& Stolte, A., 2001, ApJL, 547, 133 
\reference Johnston, K.V., Hernquist, L., \& Bolte, M., 1996, ApJ, 465, 278
\reference Kawata, D., 1999, PASJ, 51, 931
\reference Kawata, D., \& Gibson, B.K., 2003, MNRAS, 340, 908
\reference Klypin, A., Kravtsov, A.V., Valenzuela, O. \&
 Francisco, P., 1999, ApJ, 522, 82
\reference Kuijken, K. \& Dubinski, J., 1995, MNRAS, 277, 1341
\reference Moore, B., Ghigna, S., Governato, F.,
 Lake, G., Quinn, T. \& Stadel, J., 1999, ApJ, 524, L19
\reference Murphy, E.M., Lockman, F.J. \& Savage, B.D., 1995, ApJ, 447, 642
\reference Murai, T. \& Fujimoto, M., 1980, PASJ, 32, 581
\reference Noguchi, M., 1999, ApJ, 514, 77
\reference Putman, M.E.,~et al., 1998, Nature, 394, 752 
\reference Sembach, K.R., Wakker, B.P., Savage, B.D., et~al., 2003, ApJS,
 146, 165
\reference Trentham, N.,M\"oller, O., \& Ramirez-Ruiz, E.,
2001, MNRAS, 322, 658
\reference Tripp, T.M, Wakker, B.P., Jenkins, E.B., et~al, 2003, AJ, submitted
 (astro-ph/0302534)
\reference Twarog, B.A., 1980, ApJ, 242, 242
\reference Wakker, B.P., 1991, A\&A, 250, 499 
\reference Wakker, B.P., 2001, ApJS, 136, 463
\reference Wakker, B.P., Howk, J.C., Savage, B.D., van Woerden, H.,
 Tufte, S.L., Schwarz, U.J., Benjamin, R., Reynolds, R.J.,
 Peletier, R.F. \& Kalberla, P.M.W., 1999a, Nature, 402, 388
\reference Wakker, B.P., Howk, J.C., Savage, B.D., Tufte, S.L.,
 Reynolds, R.J., van Woerden, H., Schwarz, U.J. \& Peletier, R.F., 
 1999b, ASP Conf.~Ser.~166, 26
\reference Wakker, B.P., Savage, B.D., Sembach, K.R., et~al., 2003, ApJS, 146, 1
\reference Wakker, B.P. \& van Woerden, H., 1997, ARAA, 35, 217
 
\end{document}